\documentclass[acmsmall,screen]{acmart}
\AtBeginDocument{%
  }

\setcopyright{acmlicensed}
\copyrightyear{2026}
\acmYear{2026}
\acmDOI{XXXXXXX.XXXXXXX}
\acmConference[Conference acronym 'XX]{Make sure to enter the correct
  conference title from your rights confirmation email}{June 03--05,
  2026}{Woodstock, NY}
\acmISBN{978-1-4503-XXXX-X/2026/06}

\usepackage{amsmath,amsfonts}
\usepackage{algorithmic}
\usepackage{graphicx}
\usepackage{textcomp}
\usepackage{xcolor}
\usepackage{url}

\usepackage{wrapfig}
\usepackage{enumitem}
\usepackage{natbib}
\usepackage[T1]{fontenc}
\usepackage{multirow} 
\usepackage{dblfloatfix}
\usepackage{tabularx}
\usepackage[most]{tcolorbox}
\usepackage{multirow}
\usepackage{booktabs}
\usepackage{hyperref}

\usepackage{caption}
\begin{document}


\title{Towards Explorative IRBL: Combining Semantic Retrieval with LLM-driven Iterative Code Exploration}


\author{Moumita Asad}
\affiliation{%
 \institution{University of California, Irvine}
 \country{USA}}
\email{moumitaa@uci.edu}

\author{Rafed Muhammad Yasir}
\affiliation{%
 \institution{University of California, Irvine}
 \country{USA}}
\email{ryasir@uci.edu}

\author{Sam Malek}
\affiliation{%
 \institution{University of California, Irvine}
 \country{USA}}
\email{malek@uci.edu}

\renewcommand{\shortauthors}{Asad et al.}

\begin{abstract}
Information Retrieval-based Bug Localization (IRBL) aims to identify buggy source files for a given bug report. Traditional and deep learning-based IRBL techniques often suffer from vocabulary mismatch and dependence on project-specific metadata. In contrast, recent Large Language Model (LLM)-based approaches struggle to provide appropriate context to the model: they either restrict analysis to a fixed set of candidate files, overwhelm the model with repository-wide information, or rely on explicit bug report cues to guide context collection. To address these issues, we propose GenLoc, a technique that combines semantic retrieval with LLM-driven code-exploration functions to iteratively analyze the code base and identify buggy files. We evaluate GenLoc on three complementary benchmarks, including large-scale and recent Java datasets as well as the Python based SWE-bench Lite dataset. Results demonstrate that GenLoc substantially outperforms traditional IRBL, deep learning-based approaches and recent LLM-based methods, while also localizing bugs that other techniques fail to detect.
\end{abstract}




\maketitle

\section{Introduction}
Bug localization is the process of identifying the faulty or buggy locations in source code \cite{lee2018bench4bl}. It is a crucial step in software debugging to ensure the quality of a software \cite{wong2023handbook}. However, localizing bugs in practice is challenging, as large systems often incur a huge number of bugs everyday. For example, Mozilla receives around 300 new bug reports daily, which is overwhelming for the developers \cite{anvik2006should}. Moreover, given the size and complexity of large-scale software systems today, manually locating bugs is a time-consuming, tedious and an expensive task \cite{7390282}. Consequently, there is a strong demand for automated bug localization techniques to ease the burden of developers and accelerate the debugging process \cite{qi2021dreamloc}.

Over the years, researchers have proposed multiple approaches for bug localization. Among these, Spectrum-based Bug Localization (SBBL) and Information Retrieval-based Bug Localization (IRBL) are the most studied and widely recognized techniques \cite{liu2024information,qi2021dreamloc}. SBBL leverages runtime behavior (e.g., test coverage information) to analyze which parts of a program are exercised by passing and failing test cases, and ranks program elements based on their likelihood of being faulty \cite{zou2019empirical}. In contrast, IRBL takes a bug report (a document that describes a software bug) as input and does not require bug-revealing test cases or execution traces \cite{kim2021datasets}. Due to its simplicity, IR-based approaches have received growing attention in recent years.

IRBL approaches consider the bug report as a query and the software repository as a collection of documents (i.e., source files). Given a bug report, these approaches output a ranked list of source files based on their probability of being faulty. The goal is to rank the actual buggy files as high as possible. To calculate the similarity between bug reports and the source files, earlier techniques use Vector Space Model (VSM) \cite{zhou2012should,thomas2013impact}, Topic Modeling \cite{lukins2010bug,nguyen2011topic}, while later methods \cite{ye2016word,liu2019mapping,lam2015combining,lam2017bug,xiao2019improving,zhang2020exploiting,qi2021dreamloc,liang2022modeling,ye2015mapping,chakraborty2024rlocator} rely on machine learning and deep learning models. Many of these approaches also incorporate additional information such as historical bug-fix data. However, such information is not always available, particularly in new projects, limiting the generalizability of these methods. Furthermore, the effectiveness of these approaches is often limited by the vocabulary gap between bug reports and source code, as they tend to use different terminologies \cite{liu2024information}.

To bridge these limitations, recent studies \cite{Brain2025,li2025llm} have incorporated Large Language Models (LLMs) into IRBL. However, these approaches struggle with providing appropriate context to the model. These techniques (BRaIn \cite{Brain2025}, LLM-BL \cite{li2025llm}) rely on an initial retrieval step that produces a fixed shortlist of candidate files and constrain the LLM to rank files solely within this subset. Consequently, the model cannot iteratively explore the code base or gather additional evidence, limiting its effectiveness when relevant context is missing. A closely related challenge also arises in LLM-based issue localization approaches. Although these methods aim to identify fine-grained program elements (e.g., function or statement), they rely on file-level localization as an intermediate step and thus inherit similar difficulties in providing appropriate context to the LLM. Some techniques (CoSIL \cite{jiang2025cosil}, BugCerberus \cite{chang2025bridging}) expose the model to large portions of the repository, overwhelming it with irrelevant information, while others (OrcaLoca \cite{yu2025orcaloca}, LocAgent \cite{chen2025locagent}, RepoSearcher \cite{ma2508tool}) depend heavily on explicit lexical cues in the bug report and perform poorly when such signals are weak. These limitations highlight the need for a technique that can handle lexically weak bug reports, reason over a small and relevant portion of a large code base without overwhelming the LLM’s context and adaptively expand the search space by acquiring additional evidence.

In this context, we propose GenLoc (Generative AI-based Bug Localization), a novel IRBL technique that combines semantic retrieval with LLM-driven iterative code exploration to identify buggy files. Unlike many prior approaches \cite{zhou2012should,qi2021dreamloc}, GenLoc relies solely on the bug report and does not require any additional project-specific metadata or historical bug-fix data. At first, GenLoc embeds source code and retrieves semantically relevant files for a given bug report. Next, it leverages an LLM to analyze a bug report and further inspect these retrieved files, or explore additional files as needed. To enable this exploration, GenLoc provides the LLM with a set of functions (e.g., retrieving method signatures or method bodies) that support iterative analysis about which files are likely related to the reported bug. By combining semantic retrieval with LLM-guided code exploration, GenLoc reduces reliance on lexical cues, avoids the limitation of fixed candidate sets that may exclude the faulty file and prevents overwhelming the LLM with repository-wide context.

We evaluate GenLoc on three complementary benchmarks: two Java based datasets, including a large-scale benchmark of 9,097 bugs \cite{ye2015mapping} and a recent GHRB (GitHub Recent Bugs) dataset designed for LLM-based IRBL evaluation \cite{lee2024github} as well as a widely used Python benchmark (SWE-bench Lite) adopted by modern issue localization pipelines \cite{swe-bench-lite}. Results show that GenLoc consistently outperforms traditional IRBL techniques, deep learning-based approaches and recent LLM-based baselines. On the large-scale benchmark, GenLoc improves Accuracy@1 by \textbf{at least 63\%} compared to traditional and deep learning-based IRBL techniques. On the GHRB dataset, it outperforms state-of-the-art LLM-based IRBL methods by \textbf{at least 25\%}. On SWE-bench Lite, GenLoc achieves \textbf{at least 15\%} higher Accuracy@1 than recent LLM-based issue localization approaches when comparing their file-level localization stages. Furthermore, ablation studies show that GenLoc’s performance gains arise from the complementary interaction between semantic retrieval and LLM-driven iterative code exploration.

In summary, this paper makes the following contributions: 
\begin{itemize}
    \item The introduction of a novel IRBL approach named GenLoc that integrates semantic retrieval with LLM-guided iterative code analysis.
    \item A comprehensive empirical evaluation of GenLoc on three complementary benchmarks, spanning large-scale and recent Java datasets as well as a widely used Python benchmark, with comparisons against traditional IRBL baselines, LLM-based IRBL techniques and the file-level localization stage of the state-of-the-art issue localization methods.
    \item A publicly accessible replication package to facilitate future research.
\end{itemize}

\section{Background}

This section positions GenLoc within the existing literature by clarifying the problem settings, assumptions and objectives of closely related research areas. Table~\ref{taxonomy} provides a high-level comparison of these areas (Spectrum-Based Bug Localization, Information Retrieval-Based Bug Localization, Issue Localization and Issue Resolution), which are discussed in detail in Sections~2.1–2.4. Section~2.5 then positions GenLoc within this landscape and Section~2.6 introduces the terminology used throughout the paper.

\subsection{Spectrum Based Bug Localization (SBBL)}
SBBL techniques leverage program execution information collected from passing and failing test cases to localize suspicious statements or lines \cite{murali2021industry,sarhan2022survey}. Recent work, including AutoFL \cite{kang2024quantitative}, LLM4FL \cite{rafi2024enhancing}, CosFL \cite{qin2024fault}, SoapFL \cite{qin2025s}, FlexFL \cite{xu2025flexfl} and FaR-Loc \cite{shi2025enhancing}, has explored the use of LLMs in this setting. However, SBBL approaches assume the availability of failing test cases and require the faulty code be executed by at least one failing test. These assumptions limit its applicability when such tests are absent \cite{rafi2026sbest}.

\subsection{Information Retrieval Based Bug Localization (IRBL)}
IRBL techniques aim to identify buggy source files based on bug reports. Unlike SBBL, IRBL has no access to execution traces or failing test cases to constrain the search space. Thus, it needs to search across the entire code base, making the task inherently more challenging. Existing IRBL approaches can be broadly divided into three categories based on their underlying methodology. The first category \cite{lukins2008source,nguyen2011topic,zhou2012should,sisman2012incorporating,saha2013improving,moreno2014use,wang2014version,wong2014boosting,wen2016locus,wang2016amalgam+,youm2017improved,khatiwada2018just,rath2018analyzing} uses \textbf{traditional IR models} such as Latent Dirichlet Allocation (LDA) and VSM for bug localization. The second category  \cite{kim2013should,ye2014learning,fejzer2021tracking,lam2015combining,lam2017bug,xiao2019improving,koyuncu2019d,wang2020multi,cao2020bugpecker,qi2021dreamloc,iqbal2020determining,pradel2020scaffle,zhang2020exploiting,zakari2020multiple,anh2021imbalanced,liang2022modeling,luo2022improving,ciborowska2022fast,du2023pre,xiao2023bugradar,chakraborty2024rlocator,chakraborty2025blaze,li2025knowledge,zhou2026multi} consists of \textbf{learning-based models} (i.e., machine learning and deep learning models). The third category consists of \textbf{reasoning-enhanced approaches} \cite{Brain2025,li2025llm} where LLMs are used as a reasoning component to analyze and rerank retrieved candidate files. However, these methods rely on an initial retrieval step that produces a fixed shortlist of candidate files and restrict the LLM to reranking within this pre-retrieved context. Since the LLM cannot iteratively explore the code base or collect additional evidence, their effectiveness is limited.

\subsection{Issue Localization}
Issue localization focuses on identifying fine-grained code locations (e.g., methods or statements) relevant to a reported issue \cite{jiang2025agentic}. Unlike IRBL and SBBL, issue localization is not restricted to bug reports and considers heterogeneous issue types, including feature requests, performance optimizations and refactoring tasks. Although the ultimate goal of issue localization is fine-grained identification, locating relevant source files is a necessary intermediate step in all issue localization pipelines and largely determines the subsequent reasoning process. From the perspective of how relevant files are identified during localization, existing approaches in this area mainly fall into two categories. The first category consists of \textbf{iterative-reasoning methods} \cite{yu2025orcaloca,chen2025locagent,ma2508tool}, which employ an LLM to iteratively navigate the code base using lexical cues from the issue report, such as file or method names. The second category includes \textbf{repository-wide exploration methods} \cite{chang2025bridging,jiang2025cosil}, which provide the LLM with broad, repository-level context (e.g., the entire repository structure).

\subsection{Issue Resolution}
Issue resolution focuses on fixing reported issues by generating patches. MAGIS \cite{tao2024magis}, SWE-Agent \cite{yang2024swe}, OpenHands \cite{wang2024openhands}, RepoGraph \cite{ouyang2024repograph}, AutoCodeRover \cite{zhang2024autocoderover}, SpecRover \cite{ruan2024specrover}, LingmaAgent \cite{ma2025alibaba}, Agentless \cite{xia2025agentless} and SWE-Debate \cite{li2025swe} are some representative techniques in this domain. These approaches primarily focus on patch generation and test-based validation, while often overlooking the bug localization, which remains a crucial and challenging part of the debugging process \cite{chang2025bridging}.

\begin{table*}[t]
\centering
\footnotesize
\setlength{\tabcolsep}{6pt}
\renewcommand{\arraystretch}{1.2}
\caption{Comparison of SBBL, IRBL, Issue Localization and Issue Resolution.}
\begin{tabular}{|p{1.6cm}|p{2.2cm}|p{1.8cm}|p{2.7cm}|p{3.2cm}|}
\hline
\textbf{Dimension} 
& \textbf{SBBL} 
& \textbf{IRBL} 
& \textbf{Issue Localization} 
& \textbf{Issue Resolution} \\ \hline

Goal 
& Focuses on identifying buggy statements/lines 
& Focuses on identifying buggy files 
& Identifies fine-grained change locations (functions or statements) 
& Fixes reported issues \\ \hline

Input 
& Test cases and source code 
& Bug report and source code 
& Issue report and source code 
& Issue report and source code (optionally test cases) \\ \hline

Issue Types 
& Bugs only 
& Bugs only 
& Mixed issues including bugs, feature requests, and refactoring 
& Mixed issues including bugs, feature requests, and refactoring \\ \hline

Output 
& Ranked list of statements/lines 
& Ranked list of files 
& Relevant code locations 
& Patch \\ \hline

Difference from IRBL
& Requires execution traces and failing test cases, whereas IRBL relies solely on bug report
& -- 
& Targets heterogeneous issues and fine-grained code elements, using file localization only as an intermediate step
& Primarily focus on patch generation and test-based validation, while often overlooking the bug localization \cite{chang2025bridging}\\
\hline

Representative LLM-based Techniques 
& AutoFL \cite{kang2024quantitative}, LLM4FL \cite{rafi2024enhancing}, CosFL \cite{qin2024fault}, SoapFL \cite{qin2025s}, FlexFL \cite{xu2025flexfl}, FaR-Loc \cite{shi2025enhancing} 
& BRaIn \cite{Brain2025}, LLM-BL \cite{li2025llm} 
& CoSIL \cite{jiang2025cosil}, BugCerberus \cite{chang2025bridging}, OrcaLoca \cite{yu2025orcaloca}, LocAgent \cite{chen2025locagent}, RepoSearcher \cite{ma2508tool}
& MAGIS \cite{tao2024magis}, SWE-Agent \cite{yang2024swe}, OpenHands \cite{wang2024openhands}, RepoGraph \cite{ouyang2024repograph}, AutoCodeRover \cite{zhang2024autocoderover}, SpecRover \cite{ruan2024specrover}, LingmaAgent \cite{ma2025alibaba}, Agentless \cite{xia2025agentless}, SWE-Debate \cite{li2025swe} \\ \hline

\end{tabular}
\label{taxonomy}
\end{table*}

\subsection{Position of GenLoc and Motivating Example}
GenLoc addresses the problem of bug localization under the Information Retrieval-based Bug Localization (IRBL) setting. It operates exclusively on textual information from bug reports and source code and does not assume the availability of execution traces or failing test cases, distinguishing it from SBBL. Moreover, GenLoc differs in both scope and objective from issue localization and issue resolution approaches: it focuses specifically on identifying buggy source files for bug reports, rather than locating fine-grained code elements for heterogeneous issues or generating patches.

As discussed earlier, existing LLM-based IRBL approaches \cite{Brain2025,li2025llm} often fail to retrieve sufficient relevant context to support effective reasoning. Their reliance on a fixed initial shortlist prevents the LLM from incrementally expanding the search space or validating hypotheses through targeted code inspection, ultimately leading to incorrect localization results. For the bug shown in Fig.~\ref{rocketmq} from the Apache RocketMQ project\footnote{\url{https://github.com/apache/rocketmq/issues/7026}}, the correct buggy file is not included in the initial candidate shortlist produced by these approaches; consequently, the LLM has no opportunity to rerank it, making successful localization impossible.

Similarly, recent issue localization approaches \cite{yu2025orcaloca,chen2025locagent,ma2508tool,chang2025bridging,jiang2025cosil} face complementary limitations when applied to IRBL. Iterative-reasoning methods struggle with semantically rich but lexically weak issue reports, where little explicit information is available to reliably steer exploration. In contrast, repository-wide exploration methods may overwhelm the LLM by providing large context. Fig.~\ref{pylint} shows an example from the Pylint project\footnote{\url{https://github.com/pylint-dev/pylint/issues/4444}}, in which the bug report does not mention specific files, classes or methods that can be used by iterative-reasoning methods to guide the search process. In this case, repository-wide exploration methods expose the LLM to the directory structure of 778 files, while only a single file is relevant to the reported bug. This severe imbalance between relevant and irrelevant context dilutes useful signals and degrades reasoning quality. As repository size increases, the amount of extraneous context grows accordingly, further exacerbating this challenge.

These limitations highlight the need for an IRBL technique that can (1) handle semantically rich but lexically weak bug reports, (2) reason over a small, relevant portion of a large code base without flooding the LLM’s context, and (3) adaptively expand the search space by acquiring new evidence. GenLoc is designed to meet these requirements by integrating semantic retrieval with LLM-driven iterative code exploration, enabling adaptive evidence acquisition and search-space refinement during file-level localization.

\begin{figure}[hbt]
\centering
\begin{tcolorbox}[
  top=2pt,
  bottom=2pt,
  left=2pt,
  right=2pt,
  colback=gray!10!white, 
  colframe=black!75!black, 
]
\raggedright
\scriptsize{
\textbf{Bug ID:} 7027 \\
\textbf{Summary:} [Bug] The proxy in the cluster mode returns null adress when master is down\\
\textbf{Description:} When I connect to the clustered proxy to consume messages using the remoting protocol, the master broker happened to hang up. At this time, the address returned by the proxy was empty, which caused the request to be sent to the NameServer. Steps to Reproduce 1. Deploy tow group of brokers, such as: 1. master: broker-a, slave: broker-a-s 2. master: broker-b, slave: broker-b-s 2. Deploy a proxy 3. Create some topics in broker-a for testing, such as message producing and consuming 4. **Kill master broker-a**.}
\end{tcolorbox}
\caption{Bug Report from Apache RocketMQ Project.}
\label{rocketmq}
\end{figure}

\begin{figure}[hbt]
\centering
\begin{tcolorbox}[
  top=2pt,
  bottom=2pt,
  left=2pt,
  right=2pt,
  colback=gray!10!white, 
  colframe=black!75!black, 
]
\raggedright
\scriptsize{
\textbf{Bug ID:} 7114 \\
\textbf{Summary:} Linting fails if module contains module of the same name\\
\textbf{Description:}
Current behavior:
Running \texttt{pylint a} if \texttt{a/a.py} is present fails while searching for an \texttt{\_\_init\_\_.py} file. 

Expected behavior:
Running \texttt{pylint a} if \texttt{a/a.py} is present should succeed.
}
\end{tcolorbox}
\caption{Bug Report from Pylint Project.}
\label{pylint}
\end{figure}

\subsection{Terminologies}
This section introduces the key technical terms used in GenLoc.

\smallskip

\noindent \textbf{Embedding and Semantic Retrieval:}
Embedding refers to representing complex data, e.g., text, images or source code, as numerical vectors in a high-dimensional space. The core idea is that items with similar meaning (e.g., two similar sentences or two code snippets implementing similar logic) will have embeddings that are close to each other in that space. This property enables semantic retrieval, which retrieves results based on conceptual similarity rather than relying only on shared keywords or vocabulary overlap. In GenLoc, both source files and bug reports are converted into embeddings, allowing their similarity to be measured at a semantic level.

\smallskip

\noindent \textbf{Vector Database:}
A vector database is a special type of database designed to store and manage large collections of high-dimensional embeddings and perform efficient similarity searches \cite{han2023comprehensive}. To accelerate the retrieval process, vector databases commonly use Approximate Nearest Neighbor (ANN) algorithms \cite{tian2023approximate}, which find vectors that are close to a given query without exhaustively comparing all entries. In GenLoc, the embeddings of all the source files are stored in a vector database to facilitate semantic retrieval.

\smallskip

\noindent \textbf{Function Calling:} This mechanism allows LLMs to interact with external tools and systems to retrieve up-to-date information or perform specific tasks that they cannot directly do. For example, if LLM is given the prompt "\textit{What is the current temperature in Los Angeles?}", it will fail to answer because it lacks access to real-time data. However, it can be provided with access to a function like {\small\texttt{get\_current\_temperature()}}, which queries an external weather API and returns the latest temperature. The model can then call this function, receive the temperature and integrate it into its response. In GenLoc, we design a set of domain-specific code exploration functions (e.g., retrieving method signatures and method bodies) that enable the LLM to iteratively navigate the code base and identify potentially buggy files.

\smallskip

\noindent \textbf{ReAct Framework:} The ReAct (Reasoning and Acting) framework enables language models to interleave natural language reasoning with tool-based actions, thereby enhancing their problem-solving capabilities \cite{yao2023react}. Instead of generating a final answer in a single step, the model alternates between articulating intermediate reasoning steps and performing actions, such as calling a function. This synergy between reasoning and acting allows the model to iteratively refine its understanding, validate hypotheses, and make informed decisions. Since ReAct has shown promise in handling complex tasks that require multi-step thinking, access to external environments, or dynamic decision-making \cite{yao2023react,wu2025autono,li2024only}, this paper adopts the ReAct framework to guide the LLM in navigating the code base and progressively narrowing down the search space during bug localization.

\section{Methodology}
\begin{figure}[t]
\centering
\includegraphics[width=\textwidth]{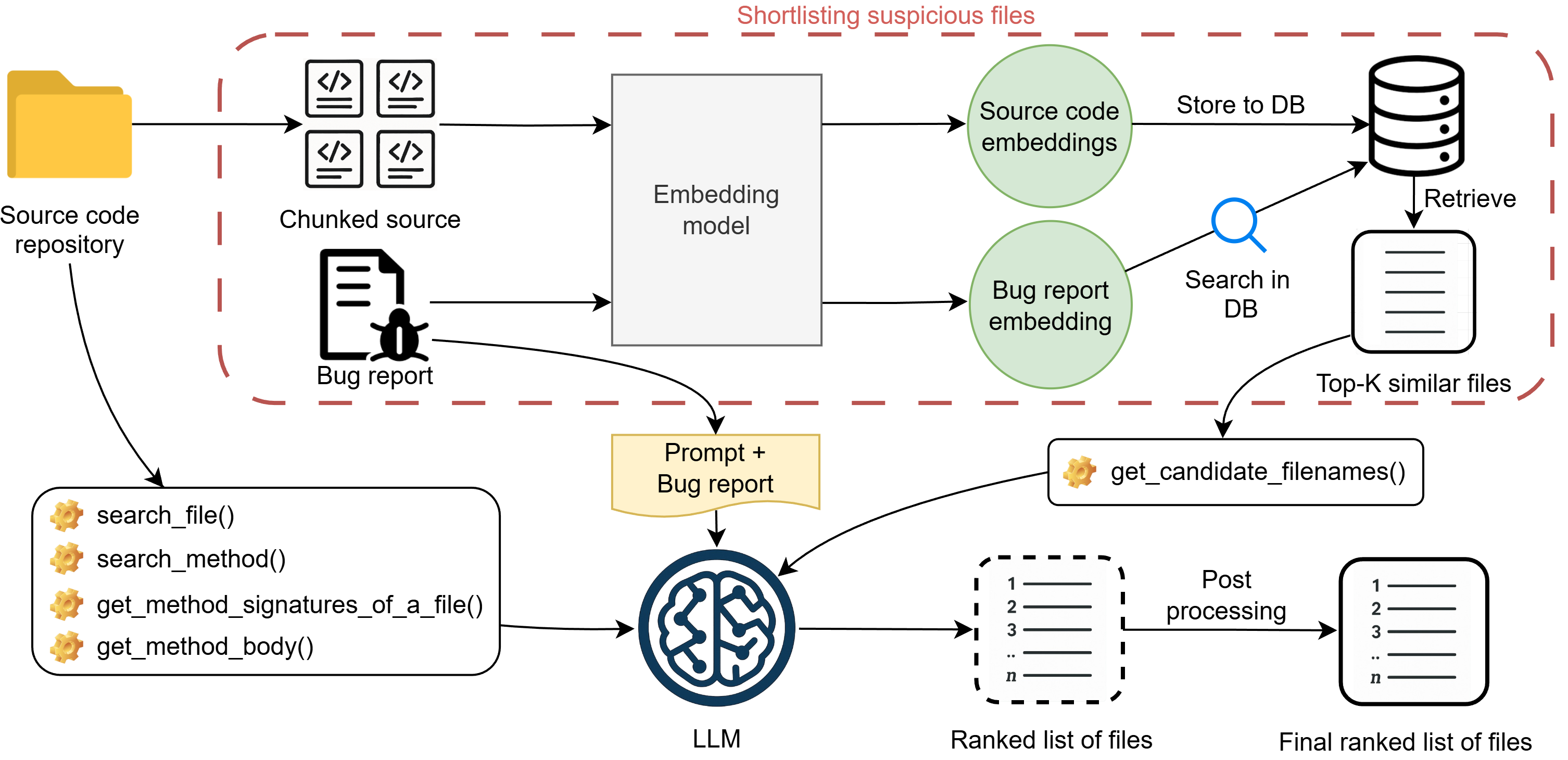}
\caption{Workflow of GenLoc.}
\label{workflow}
\end{figure}

GenLoc operates in two primary steps to localize relevant files based on a given bug report. First, it retrieves a set of semantically similar files using embedding-based similarity. Next, it employs an LLM augmented with a set of custom-designed code exploration functions, which allow the model to iteratively reason over the bug report and interact with the code base. During this stage, the model may examine the embedding-based retrieved files or explore other parts of the code base to generate a ranked list of potentially buggy files. An overview of the workflow is shown in Fig. \ref{workflow}.

\subsection{Shortlisting suspicious files}\label{step_a}
To identify source files potentially related to a given bug report, this step computes the semantic similarity between the bug report and the source code using embedding-based representations. The process begins by traversing all source files in the project. For each file, its fully qualified file path and method bodies are extracted. Each file is then represented as a concatenation of its fully qualified file path and the method bodies. The inclusion of the file path provides contextual information about the file’s role or feature within the overall system architecture (e.g., {\footnotesize\path{org.eclipse.jdt.ui/ui/org/eclipse/jdt/ui/JavaElementLabels.java}} suggests a UI-related component), while method bodies capture the actual implementation logic, which is crucial since methods serve as the fundamental units of behavior in a program.

The concatenation of a file's path and method bodies can result in lengthy text. To better localize relevant content, this representation is split into smaller chunks if it exceeds a predefined length threshold. Each chunk is then converted into an embedding vector and stored in a vector database. Similarly, the textual content of the bug report, comprising its summary and description, is transformed into embedding vector. The bug report embedding is then used to query the vector database, retrieving the most semantically similar file chunks. Based on these retrieval results, a predefined number of top-ranked files are shortlisted as suspicious. These files serve as candidate inputs for further analysis by the LLM in the subsequent step.


\subsection{Ranking suspicious files}
In this step, an LLM is employed to generate a ranked list of potentially buggy source files based on a given bug report, as illustrated in Fig. \ref{workflow}. To support the LLM’s code exploration process, a prompt (shown in Fig. \ref{prompt}) is constructed by following the ReAct (Reasoning and Acting) framework \cite{yao2023react}, which enables LLMs to interleave natural language reasoning with tool-based actions. This approach has proven effective in improving performance on complex decision-making tasks by enabling iterative reasoning and environment interaction.

\begin{figure*}[!ht]
\centering
\begin{tcolorbox}[
  top=2pt,
  bottom=2pt,
  left=4pt,
  right=4pt,
  colback=gray!10!white, 
  colframe=black!75!black, 
  width=\textwidth,
]

\scriptsize 

You are an expert software engineer specializing in bug localization. Your goal is to identify the most probable buggy Java files based on a given bug report. You have access to five functions that help you search for file names, locate methods and analyze source code. You must follow an iterative, reasoning-based approach, refining your strategy dynamically based on insights gathered during each step.

At each iteration, you should:
\begin{itemize}[leftmargin=2em]
\item Maintain a working shortlist of files that appear potentially buggy.
\item Update the shortlist based on new evidence (e.g., method matches, code analysis).
\item Avoid redundant operations--do not recheck the same filenames, methods or file contents multiple times.
\end{itemize}

Continue this process until you either: (a) Produce a well-justified ranked list of the 10 most relevant files, or (b) Reach the maximum limit of 10 iterations. In the 10th iteration, you must output your final ranked list.

Workflow

\textbf{1. Analyze the Bug Report:}
    \begin{itemize}[leftmargin=2em]
        \item Extract relevant keywords, error messages and functional hints from the bug summary and description.
        \item Identify potentially affected components (e.g., UI, database, networking).
    \end{itemize}
    
\textbf{2. File Discovery:}
    \begin{itemize}[leftmargin=2em]
        \item Use \textit{search\_file()} to check if filenames derived from the bug report's keywords or functionality exist in the code base.
        \item If the bug report references a specific method name, use \textit{search\_method()} to find all files defining it.
        \item If an inferred filename or method location does not exist, refine your strategy: adjust assumptions, explore variations and retry.
        \item If strong matches are not found, use \textit{get\_candidate\_filenames()} to retrieve 50 potentially relevant filepaths.
        \item Add promising files to your shortlist by prioritizing those align with terminology, functionality or methods discussed in the bug report.
    \end{itemize}
    
\textbf{3. Method Analysis:}
    \begin{itemize}[leftmargin=2em]
        \item For each file in the shortlist that hasn't been analyzed yet:
        \begin{itemize}[leftmargin=1em]
            \item Use \textit{get\_method\_signatures\_of\_a\_file()} to list its methods.
            \item Identify methods that directly align with the bug's context (e.g., related functionality, naming hints).
            \item For any method of interest, retrieve its implementation using \textit{get\_method\_body()}.
            \item Analyze the logic of any identified method(s) of interest to determine whether they align with the bug's symptoms.
        \end{itemize}
        \item If this analysis reveals new relevant class names, filenames, or methods, use the appropriate search functions.
        \item Continuously update the shortlist by promoting, demoting or removing files based on evolving understanding.
    \end{itemize}

\textbf{4. Shortlist Refinement \& Ranking:}
    \begin{itemize}[leftmargin=2em]
        \item Rank files based on:
        \begin{itemize}
            \item Semantic alignment with the bug report's keywords and described functionality
            \item Method or filename alignment with bug context
            \item Code logic alignment with the bug description
        \end{itemize}
        \item If needed, iterate with refined assumptions or explore previously overlooked filenames or methods.
    \end{itemize}

\textbf{5. Final Output:}
    \begin{itemize}[leftmargin=2em]
        \item Provide a ranked list of the 10 most relevant filepaths based on their likelihood of containing the bug.
        \item Ensure filepaths exactly match those provided--do not modify case, structure or abbreviate them.
        \item Justify each file's inclusion by referencing keywords, method matches or code logic that supports its relevance to the bug report.
    \end{itemize}

\end{tcolorbox}
\caption{LLM Prompt.}
\label{prompt}
\end{figure*}

At first, the LLM is prompted to assume the role of an expert software engineer specializing in bug localization, as role-playing improves the reasoning abilities of LLMs \cite{kong2023better}. Next, the prompt is designed to decompose bug localization into a series of smaller tasks since LLMs perform better when complex tasks are broken down into sub-tasks \cite{khot2022decomposed}. These sub-tasks include analyzing the bug report, identifying potential file or method names, analyzing method signatures and bodies and integrating evidence to prioritize suspicious files. 

To support the analysis process, GenLoc provides the LLM with access to the following five external functions. The LLM may invoke one or more of these functions over a fixed number of iterations to get information about the code base. Collectively, these functions allow the LLM to discover files, understand their role and analyze the internal logic of specific code segments, thereby supporting both breadth and depth in the bug localization process.

\begin{enumerate}[leftmargin=*]
\item {\small\texttt{search\_file()}}: This function enables the LLM to check whether a specific file exists in the code base. It can be used to (i) verify the existence of a file explicitly mentioned in the bug report (e.g., checking if {\small\texttt{JavaElementLabels.java}} exists when “JavaElementLabels” is referenced), (ii) infer and validate a possible filename derived from textual clues in the bug report, or (iii) identify candidate files after inspecting a method body.

\item {\small\texttt{search\_method()}}: When a method name is mentioned in the bug report, e.g., {\small\texttt{updateLabel()}}, this function enables the LLM to find which files contain its definition and thereby narrowing down the search space. In addition, it can be employed to discover candidate files based on references observed during method body inspection.

\item {\small\texttt{get\_candidate\_filenames()}}: This function retrieves a list of potentially relevant files based on their semantic similarity to the bug report from the previous step (Section \ref{step_a}). The LLM can leverage this list as an initial pool of candidates, especially when the bug report does not provide any clues.

\item {\small\texttt{get\_method\_signatures\_of\_a\_file()}}: Once a file is identified, this function enables the model to inspect the file’s high-level structure by retrieving all method signatures defined within it. This helps the LLM determine whether the file's responsibilities align with the bug report.

\item {\small\texttt{get\_method\_body()}}: To evaluate a method's implementation, the model can call this function to retrieve the body of a given method. This allows a deeper inspection of whether the method may be related to the described issue.
\end{enumerate}


At each iteration, the model can “reason” about what information is currently available, determine what additional data is required, and then “act” by invoking one of these functions. The outcome of the function call is integrated into the LLM’s internal reasoning in the subsequent iteration. Since function calls issued by the LLM may contain errors, such as supplying incorrect parameters like non-existent filenames or method signatures, GenLoc incorporates an iteration-time recovery step. When no exact match is found, GenLoc either provides tentative options (e.g., all file paths sharing the same base filename or the closest method signatures using Damerau–Levenshtein distance \cite{zhao2019string}) or returns explicit feedback indicating that the requested element does not exist in the code base. This mechanism guides the LLM in handling mistakes (e.g, typos or incomplete names) by recovering the intended file or method, or preventing it from producing non-existent ones.

Lastly, the LLM is prompted to output a ranked list of top 10 files, where each entry includes a fully qualified file path along with a brief justification for its selection. This design choice is informed by empirical findings indicating that inspecting more than ten files exceeds the acceptability threshold for nearly 98\% of practitioners \cite{kochhar2016practitioners}. Since multiple files from different packages can have the same name, the fully qualified file paths are used to uniquely identify each file. The inclusion of justifications is motivated by prior work demonstrating that self-explanatory prompts significantly improve LLM comprehension \cite{wu2023large}.

Although the LLM receives feedback during intermediate iterations, it can still produce erroneous or non-existent file paths in the final ranked list. A common issue observed in our preliminary analysis is the omission of intermediate packages. For example, the model may omit the "internal" package and produce {\footnotesize\path{org.eclipse.jdt.ui/ui/org/eclipse/jdt/ui/text/java/AlphabeticSorter.java}} instead of
{\footnotesize\path{org.eclipse.jdt.ui/ui/org/eclipse/jdt/internal/ui/text/java/AlphabeticSorter.java}}. To address this, GenLoc applies a post-processing recovery step to verify the correctness of each predicted file. If the predicted fully qualified file path matches an existing file, it is retained. Otherwise, the base filename (e.g., {\small\texttt{AlphabeticSorter.java}}) is extracted, and all files with the same name are retrieved. Among these candidates, the file whose fully qualified file path has the highest Jaccard similarity \cite{niwattanakul2013using} with the predicted filename is selected. Since Jaccard similarity measures token overlap, it favors the correct file even when an intermediate package is missing. If the base filename does not exist in the code base, no suitable match can be found and the file is excluded from the final ranked list.

\section{Experiement Setup}
This section presents the implementation details, research questions, baseline techniques, benchmark dataset and evaluation metrics.

\subsection{Research Questions}
To evaluate GenLoc, the following research questions are investigated:

\begin{itemize}[label={}, leftmargin=0pt]
    \item \textbf{RQ1:} How does GenLoc perform compared to traditional and deep learning-based IRBL techniques?
    \item \textbf{RQ2:} How does GenLoc perform compared to other LLM-based IRBL techniques?
    \item \textbf{RQ3:} How does GenLoc perform compared to the file-level localization stages of LLM-based issue localization techniques?
    \item \textbf{RQ4:} How do different components of GenLoc contribute to the overall performance?
    \item \textbf{RQ5:} How well does GenLoc perform on previously unseen bugs?
    \item \textbf{RQ6:} How does the choice of retrieval model and LLM impact the performance of GenLoc?
\end{itemize}

\subsection{Baseline Techniques} 
To comprehensively evaluate GenLoc, we compare it against publicly available and executable baselines from three categories.

\begin{itemize}[label={},leftmargin=0pt]
    \item \textbf{Traditional and Deep Learning-Based IRBL Techniques (RQ1).}
We select four widely studied IRBL techniques from the traditional and deep learning categories: BugLocator \cite{zhou2012should}, BLUiR \cite{saha2013improving}, BRTracer \cite{wong2014boosting} and DreamLoc \cite{qi2021dreamloc}. Although BugLocator, BLUiR and BRTracer were introduced earlier, they remain actively used in recent empirical studies to characterize the strengths and limitations of IR-based bug localization techniques \cite{li2022empirical,kim2021datasets}. BugLocator uses revised VSM (rVSM) by incorporating file length normalization and historical bug-fix frequency to prioritize more fault-prone files \cite{zhou2012should}.
BLUiR assigns different weights to structural program elements such as class and method names, and computes their similarity to the bug report content separately \cite{saha2013improving}.
BRTracer divides source files into smaller segments and integrates additional signals such as stack traces and historical fix data to rank files \cite{wong2014boosting}.
DreamLoc leverages a Wide and Deep architecture: the Wide component captures software-specific statistical features (e.g., fix frequency and recency), while the Deep component applies a relevance model for fine-grained similarity scoring \cite{qi2021dreamloc}. 

\hspace*{1.5em} For DreamLoc, the original implementation is used. For BugLocator, BLUiR and BRTracer, the implementations provided in \cite{lee2018bench4bl} are adopted, since the original implementations are unavailable. While we aimed to incorporate several other IRBL techniques in the evaluation, they could not be included due to implementation-related issues. For instance, DNNLOC \cite{lam2017bug} and DeepLoc \cite{xiao2019improving} do not have a publicly released implementation. Although RLocator \cite{chakraborty2024rlocator} provides a replication package, it is incomplete due to missing source code. AdaptiveBL \cite{fejzer2021tracking} and its successor FLIM \cite{liang2022modeling} could not be executed due to unresolved dependency issues.

\smallskip

\item \textbf{LLM-Based IRBL Techniques (RQ2).} Following the taxonomy in Table~\ref{taxonomy}, we compare GenLoc against two state-of-the-art LLM-based IRBL approaches: BRaIn \cite{Brain2025} and LLM-BL \cite{li2025llm}. BRaIn retrieves top-K files using lexical search, identifies relevant methods from those files using an LLM and refines the results through keyword-based query expansion and reranking \cite{Brain2025}. LLM-BL expands the bug report using an LLM, retrieves candidate files through hybrid (lexical and semantic) search and presents the shortlisted files along with their relevant code lines to the LLM for reranking \cite{li2025llm}. For both techniques, we use their original implementations.

\smallskip

\item \textbf{File-Level Localization Stages of Issue Localization Techniques (RQ3).} As summarized in Table~\ref{taxonomy}, issue localization techniques aim to identify fine-grained change locations but generally begin with a file-level localization phase to narrow down the search space before subsequent analysis. Accordingly, we compare GenLoc against the file-level localization stages of two state-of-the-art LLM-based issue localization approaches: CoSIL \cite{jiang2025cosil} and LocAgent \cite{chen2025locagent}. These approaches represent repository-wide exploration and iterative-reasoning methods, respectively. Prior studies show that CoSIL outperforms other issue localization techniques such as OrcaLoca as well as RepoSearcher at the file level \cite{jiang2025cosil}. We exclude BugCerberus due to the unavailability of its implementation and because it adopts the AgentLess paradigm \cite{xia2025agentless}, which is outperformed by CoSIL \cite{jiang2025cosil}.

CoSIL performs file-level localization by prompting the LLM to rank suspicious files using the repository’s directory tree and then uses a Python import-based call graph to refine the localization results \cite{jiang2025cosil}. LocAgent extracts keywords from the issue description and uses them to guide dependency-graph exploration for progressively identifying relevant files \cite{chen2025locagent}. We use the original implementations of both approaches and evaluate them with the same LLM as GenLoc to ensure a fair comparison under identical model conditions, while restricting the evaluation strictly to their file-level localization outputs.


\end{itemize}

\subsection{Benchmark Dataset} 

We use three diverse benchmark datasets to evaluate GenLoc: the Ye et al. dataset \cite{ye2015mapping}, the GHRB (GitHub Recent Bugs) dataset \cite{lee2024github} and SWE-bench Lite \cite{swe-bench-lite}. The Ye et al. dataset is the most widely adopted benchmark in prior IRBL studies \cite{qi2021dreamloc,ye2016word,fejzer2021tracking}. Additionally, its large scale, over 22,000 bugs from six large, open-source Java projects (Table \ref{dataset}), makes it well suited for training deep learning–based approaches (e.g., DreamLoc) and comparing with GenLoc \cite{ye2014learning,ye2015mapping}. The GHRB (GitHub Recent Bugs) dataset contains 131 recent bugs from 16 Java projects (e.g., Apache Dubbo, OpenAPI Generator), as shown in Table \ref{ghrb}. This dataset is specifically designed for evaluating LLM-based IRBL approaches \cite{lee2024github}. SWE-bench Lite is a curated subset of SWE-bench \cite{jimenez2023swe} that focuses mainly on bug-fixing issues \cite{swe-bench-lite}. It contains 300 issues from 12 popular Python projects, including Django, Matplotlib (Table \ref{swebenchlite}) and has been widely adopted by recent issue localization approaches, making it suitable for evaluating file-level localization in issue localization pipelines \cite{jiang2025cosil, xia2025agentless}.

\begin{table}[ht]
\footnotesize
\centering
\caption{Ye et al. Dataset Description}
\begin{tabular}{|c|c|rrr|rrr|}
\hline
\multirow{2}{*}{\textbf{Project}} & \multirow{2}{*}{\textbf{\# of Bug Reports}} & \multicolumn{3}{c|}{\textbf{\# of Java Files}} & \multicolumn{3}{c|}{\textbf{LOC}}             \\ \cline{3-8} 
                                  &                                             & \textbf{Max}  & \textbf{Median} & \textbf{Min} & \textbf{Max} & \textbf{Median} & \textbf{Min} \\ \hline
AspectJ                           & 593                                         & 6,879         & 4,439           & 2,076        & 699,250      & 515,153         & 216,387      \\ \hline
Birt                              & 4,178                                       & 9,697         & 6,841           & 1,700        & 2,322,074    & 1,549,525       & 412,795      \\ \hline
Eclipse                           & 6,495                                       & 6,243         & 3,454           & 382          & 1,156,041    & 637,158         & 47,348       \\ \hline
JDT                               & 6,274                                       & 10,544        & 8,184           & 2,294        & 922,812      & 609,378         & 133,323      \\ \hline
SWT                               & 4,151                                       & 2,795         & 2,056           & 1,037        & 934,357      & 639,216         & 301,698      \\ \hline
Tomcat                            & 1,056                                       & 2,042         & 1,552           & 924          & 487,701      & 413,472         & 270,046      \\ \hline
\end{tabular}
\label{dataset}
\end{table}



\begin{table}[ht]
\footnotesize
\centering
\caption{GHRB Dataset Description}
\begin{tabular}{|c|c|ccc|ccc|}
\hline
\multirow{2}{*}{\textbf{\# of Projects}} & \multirow{2}{*}{\textbf{\# of Bug Reports}} & \multicolumn{3}{c|}{\textbf{\# of Java Files}} & \multicolumn{3}{c|}{\textbf{LOC}}             \\ \cline{3-8} 
                                  &                                             & \textbf{Max}  & \textbf{Median} & \textbf{Min} & \textbf{Max} & \textbf{Median} & \textbf{Min} \\ \hline
16                           & 131                                         & 12,025         & 1,977           & 93        & 1,874,139      & 299,697         & 14,759      \\ \hline
\end{tabular}
\label{ghrb}
\end{table}


\begin{table}[ht]
\footnotesize
\centering
\caption{SWE-bench Lite Dataset Description}
\begin{tabular}{|c|c|ccc|ccc|}
\hline
\multirow{2}{*}{\textbf{\# of Projects}} &
\multirow{2}{*}{\textbf{\# of Bug Reports}} &
\multicolumn{3}{c|}{\textbf{\# of Python Files}} &
\multicolumn{3}{c|}{\textbf{LOC}} \\ \cline{3-8}
 &  & \textbf{Max} & \textbf{Median} & \textbf{Min} & \textbf{Max} & \textbf{Median} & \textbf{Min} \\ \hline
12 & 300 & 2,757 & 1,377 & 75 & 771,389 & 361,222 & 16,551 \\ \hline
\end{tabular}
\label{swebenchlite}
\end{table}

\subsection{Evaluation Metrics} 
Similar to other work \cite{zhou2012should,rahman2018improving}, the following three metrics are used for evaluation:

\smallskip
\noindent \textbf{1. Accuracy@k:} It denotes the number of bugs for which at least one of the actual buggy files is ranked within the top-k positions (k = 1, 5, 10) of the ranked list. A higher accuracy@k value indicates a better performance of the bug localization technique. Similar to other studies \cite{qi2021dreamloc,liang2022modeling}, the value of k is set to 1, 5 and 10.

\smallskip

\noindent \textbf{2. Mean Reciprocal Rank@k (MRR@k):} This metric measures the average reciprocal rank of the first correctly identified buggy file within the top-k results across all bug reports \cite{wong2023handbook}. For each bug report, the reciprocal rank is defined as the inverse rank of the first correct buggy file that appears among the top-k results. MRR@k is calculated as follows: 
\begin{equation}
MRR@k = \frac{1}{N} \sum_{i=1}^{N} 
\begin{cases}
\frac{1}{\text{rank}_i} & \text{if } \text{rank}_i \leq k \\
0 & \text{otherwise}
\end{cases}
\label{mrr}
\end{equation}
where, $N$ is the total number of bug reports. $\text{rank}_i$ denotes the rank of the first correct buggy file for the i-th bug report among the top-k results. The higher the value of MRR@k, the better the technique. Following prior studies \cite{rahman2018improving,kim2019novel}, the value of k is set to 10. An MRR@10 value of 0.2 indicates that, on average, the first relevant file is ranked around position 5 within the top 10 results. 

\smallskip

\noindent \textbf{3. Mean Average Precision@k (MAP@k):} This metric evaluates how well a bug localization technique ranks all actual buggy files within the top-k retrieved results for a set of bug reports \cite{hirsch2023map}. For each bug report, it calculates the Average Precision (AP) using the following formula, which considers both the presence and the ranking positions of all relevant buggy files: 

\begin{equation}
AP_i = \frac{ \sum_{j=1}^{M} P(j) \times \text{Rel}(j) }{ \text{Number of relevant source files} }
\label{ap}
\end{equation}
where, $M$ is the number of retrieved files in the top-k ranked list for the i-th bug report (i.e., $M$=$k$). $Rel(j)$ denotes whether the file at position $j$ is relevant or not. $P(j)$ indicates the precision at rank $j$ and computed as follows:

\begin{equation}
P(j) = \frac{ \text{Number of relevant files in top-} j \text{ positions} }{ j }
\label{p}
\end{equation}

The Mean Average Precision is then computed by averaging the AP scores across all bug reports, as shown here:
\begin{equation}
MAP = \frac{ \sum_{i=1}^{N} AP_i }{N}
\label{map}
\end{equation}
where, $N$ denotes the total number of bug reports. A higher MAP@k value indicates better ranking quality of the buggy files. Following prior studies \cite{rahman2018improving,kim2019novel}, the value of k is set to 10. For example, a MAP@10 value of 0.3 implies that, on average, 30\% of the actual buggy files are retrieved with high precision within the top 10 results.

\subsection{Implementation}
GenLoc currently supports Java and Python and can be easily extended to other programming languages, as it leverages the Tree-sitter library \cite{tree-sitter} for source code parsing, which supports over ten languages. For semantic retrieval, GenLoc uses OpenAI’s {\small\texttt{text-embedding-3-small}} as its default model due to its cost-effectiveness (\$0.02 per 1M tokens) \cite{text-embedding}. To further reduce embedding cost and processing time, GenLoc stores all source-code chunk embeddings in ChromaDB \cite{chroma} when processing the first bug of a project. For subsequent bugs, embeddings are updated incrementally by re-embedding only files that are added, modified, deleted or renamed across versions\footnote{GenLoc takes on average 47.7 seconds per bug in our experiments; detailed runtime and cost analyses are provided in the replication package.}. For inference, GenLoc uses OpenAI's {\small\texttt{GPT-4o-mini}} as the default LLM due to its cost-effectiveness (\$0.15 per 1M input tokens and \$0.60 per 1M output tokens) \cite{gpt-mini}, which enables large-scale experimentation on more than 9,500 bugs across three datasets. In \hyperref[sec:rq6]{RQ6}, we further evaluate the impact of using a more capable model ({\small\texttt{GPT-5.1}}) on GenLoc’s performance. Since LLM outputs are non-deterministic, all experiments are executed three times and results are reported as averages, following prior work \cite{zhang2024autocoderover}. Beyond model choices, key implementation decisions (e.g., chunk size and the number of shortlisted candidates) are determined empirically, with all configuration details provided in the replication package.

\section{Result Analysis}
In this section, the experimental results in accordance with the research questions are discussed.

\subsection{RQ1: Comparison with Traditional and Deep Learning-based IRBL Techniques}

Table \ref{overall} reports the results of evaluating GenLoc against four traditional and deep learning-based IRBL techniques (DreamLoc, BRTracer, BLUiR and BugLocator). We use the Ye et al. dataset due to its scale and popularity \cite{fejzer2021tracking,chakraborty2024rlocator,maarleveld2025gotta}. Following prior work \cite{han2023bjxnet,chakraborty2024rlocator}, 60\% data is used as historical (training) data and the most recent 40\%, consisting of 9,097 bugs, is used for evaluation.

Results show that GenLoc demonstrates strong overall performance across all evaluation metrics compared to prior bug localization techniques. On average, GenLoc achieves the highest Accuracy@1 (44.01\%), substantially outperforming the second-best technique (BRTracer with 26.84\%) by a relative margin of over 63\%. This indicates that GenLoc can correctly identify the buggy file at the top position in a significant portion of bug reports. Similarly, GenLoc leads in Accuracy@5 (63.27\%) and Accuracy@10 (68.28\%), further affirming its ability to prioritize buggy files effectively within the top candidates. In terms of ranking quality, GenLoc achieves the highest scores in both MAP@10 (0.41) and MRR@10 (0.52), outperforming the next-best MAP@10 (DreamLoc with 0.30) and MRR@10 (BRTracer and DreamLoc with 0.37) by 36.67\% and 40.54\%, respectively, confirming its superiority in ranking relevant files near the top of the result list.

\begin{table}[ht]
\footnotesize
\centering
\caption{Comparison of GenLoc and Non-LLM-based IRBL Techniques}
\begin{tabular}{cccccc}
\hline
\multirow{2}{*}{\textbf{Technique}} & \multicolumn{3}{c}{\textbf{Accuracy@k (\%)}} & \multirow{2}{*}{\textbf{MAP@10}} & \multirow{2}{*}{\textbf{MRR@10}} \\
\cline{2-4}
 & k=1 & k=5 & k=10 & & \\
\hline
GenLoc      & \textbf{44.01} & \textbf{63.27} & \textbf{68.28} & \textbf{0.41} & \textbf{0.52} \\
DreamLoc    & 26.33 & 52.32 & 63.07 & 0.30 & 0.37 \\
BRTracer    & 26.84 & 51.21 & 60.42 & 0.29 & 0.37 \\
BLUiR       & 26.49 & 49.04 & 59.82 & 0.28 & 0.36 \\
BugLocator  & 24.19 & 46.20 & 56.65 & 0.26 & 0.33 \\
\hline
\end{tabular}
\label{overall}
\end{table}


It is worth mentioning that GenLoc uses only the bug report as input, without incorporating any additional information such as historical bug-fix data or file change frequency. In contrast, techniques like DreamLoc, BRTracer and BugLocator leverage such additional data to improve localization accuracy. Despite considering less information, GenLoc consistently outperforms these baselines on all metrics.

\begin{wrapfigure}{r}{0.42\textwidth}
\vspace{-10pt} 
\centering
\includegraphics[width=\linewidth]{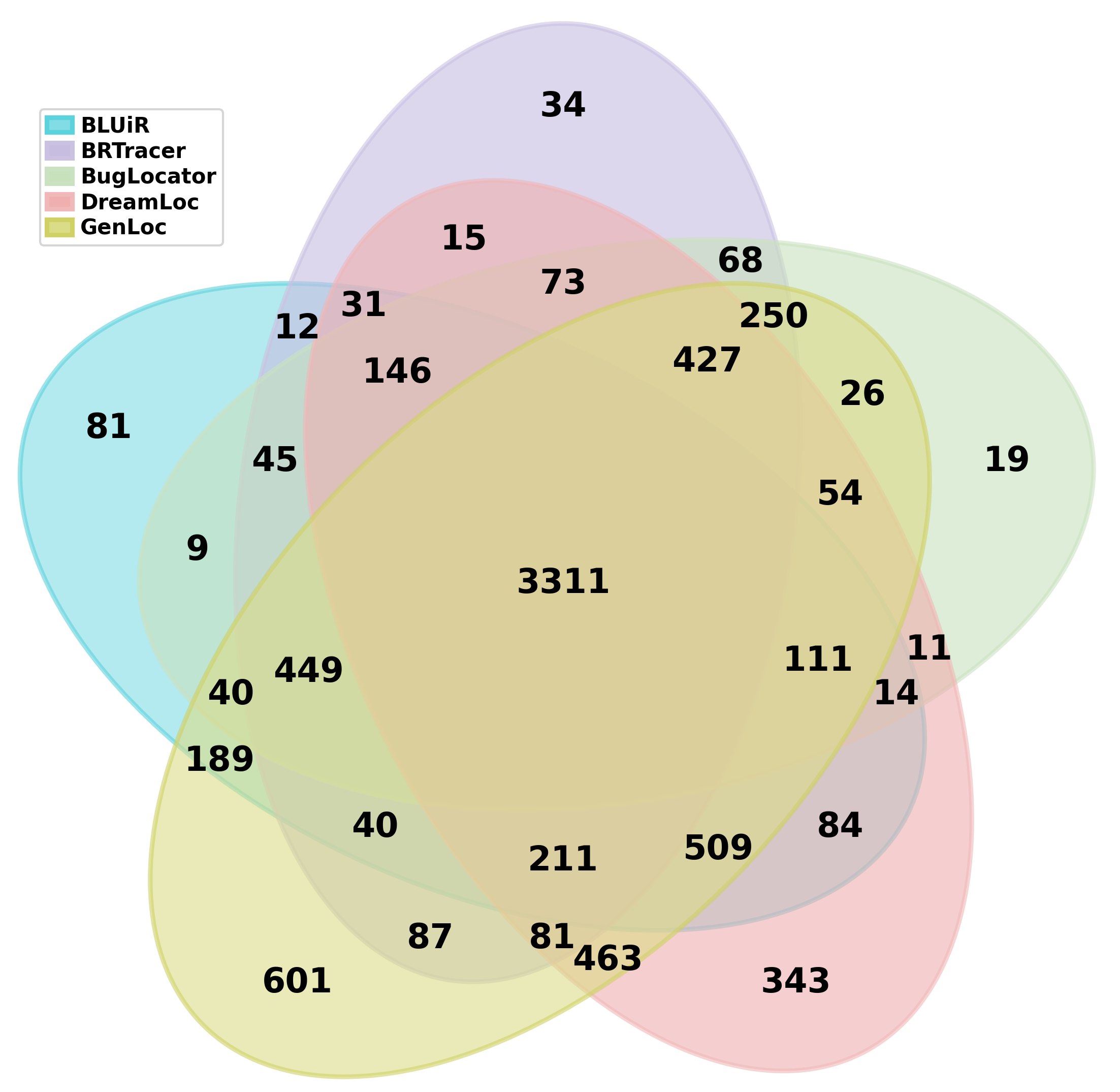}
\caption{Overlap Analysis between GenLoc and Non-LLM-based IRBL techniques.}
\label{overlap-non-llm}
\end{wrapfigure}

Apart from Accuracy@k, MRR@10 and MAP@10, the number of unique bugs localized by each approach is examined. In this analysis, a bug is considered successfully localized if the correct buggy file appears within the top 10 ranked results (i.e., Accuracy@10) \cite{kochhar2016practitioners}. Figure \ref{overlap-non-llm} presents the unique bugs localized by each technique under this criterion. To account for GenLoc’s variability across trials and enable a fair comparison with existing deterministic baselines, we consider the union of all bugs localized across three trials. This union captures the set of unique bugs that GenLoc can localize in at least one run, reflecting the overall capability of the approach rather than the outcome of a single arbitrary execution.


Among all approaches, GenLoc can localize the highest number of unique bugs (601 bugs). This indicates that GenLoc not only localizes more bugs but also captures a unique subset of bugs missed by all other methods, highlighting its potential to complement existing methods.

\subsection{RQ2: Comparison with LLM-based IRBL Techniques}

We compare GenLoc against two state-of-the-art LLM-based IRBL techniques, BRaIn and LLM-BL. We use both the GHRB dataset and a subset of the Ye et al. dataset consisting of the most recent 300 bug reports (50 per project) to enable assessment across both modern and traditional benchmarks.

\begin{table}[h]
\footnotesize
\caption{Comparison of GenLoc and LLM-based IRBL Techniques}
\begin{tabular}{cccccc}
\cline{1-6}
\multirow{2}{*}{\textbf{Technique}} & \multicolumn{3}{c}{\textbf{Accuracy@k (\%)}}                                                            & \multicolumn{1}{c}{\multirow{2}{*}{\textbf{MAP@10}}} & \multicolumn{1}{c}{\multirow{2}{*}{\textbf{MRR@10}}} \\ \cline{2-4}
                                    & \multicolumn{1}{c}{\textbf{k=1}} & \multicolumn{1}{c}{\textbf{k=5}} & \multicolumn{1}{c}{\textbf{k=10}} & \multicolumn{1}{c}{}                                 & \multicolumn{1}{c}{}                                 \\ \hline
GenLoc                              & \textbf{63.36}                   & \textbf{76.85}                   & \textbf{79.65}                    & \textbf{0.59}                                        & \textbf{0.69}                                        \\
BRaIn                               & 22.90                            & 51.15                            & 61.07                             & 0.28                                                 & 0.34                                                 \\
LLM-BL          & 50.38                            & 63.11                            & 67.18                             & 0.47                                                 & 0.56      \\
\bottomrule
\end{tabular}
\label{llm-comparison}
\end{table}

\begin{wrapfigure}{r}{0.36\textwidth}
\vspace{-13pt}
\begin{center}
\includegraphics[width=0.36\textwidth]{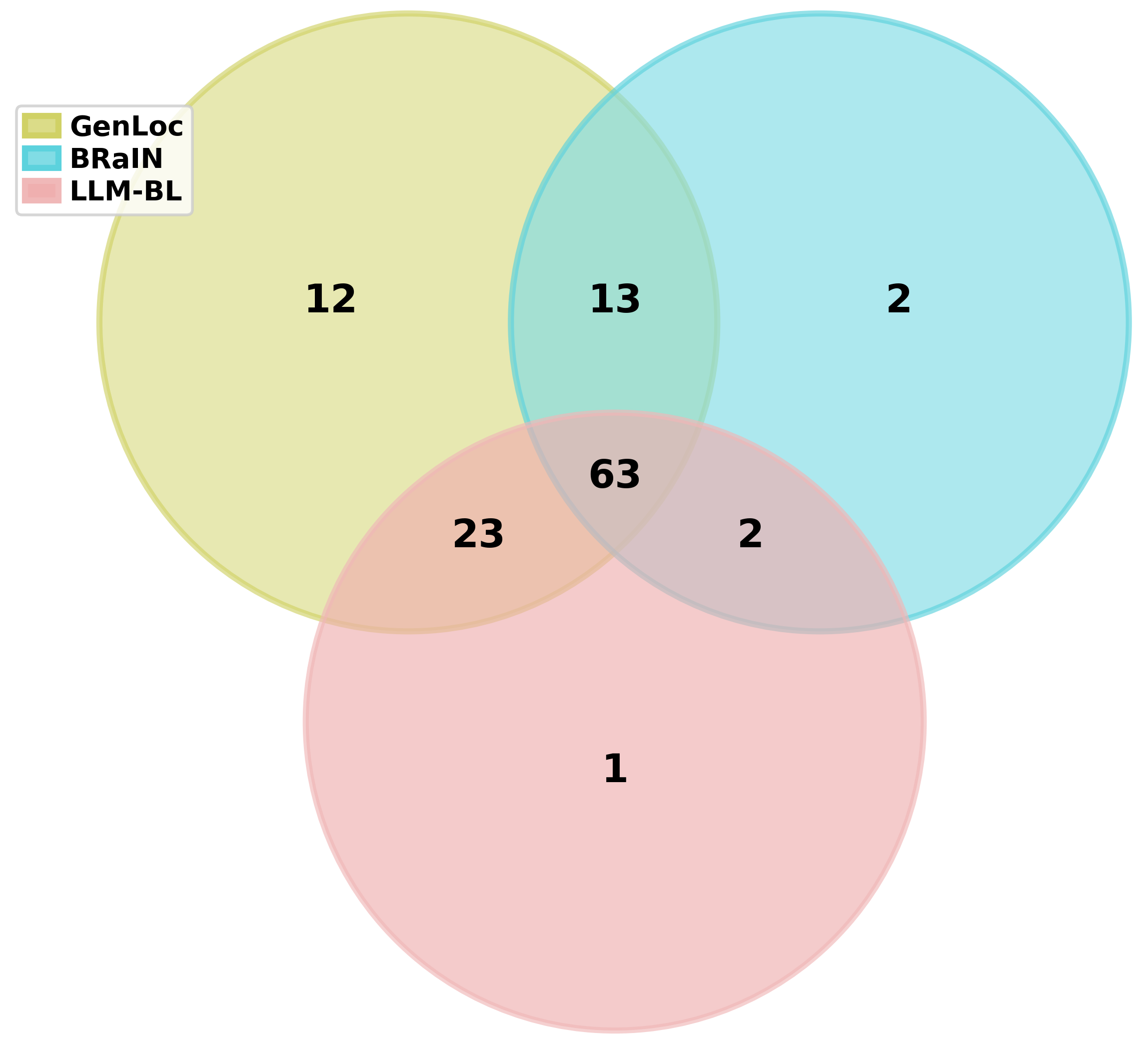}
\end{center}
\caption{Overlap Analysis between GenLoc and Other LLM-based IRBL Approaches.}
\label{overlap-llm}
\vspace{-16pt}
\end{wrapfigure}

Table \ref{llm-comparison} reports the comparison on the GHRB dataset. The results show that GenLoc consistently outperforms all baselines across every evaluation metric. It achieves the highest Accuracy@1 of 63.36\%, substantially higher than other techniques and maintains this advantage at k=5 and k=10. For ranking-based measures, GenLoc achieves MAP@10 of 0.59 and MRR@10 of 0.69, representing over 23\% improvement compared to LLM-BL and more than 100\% improvement over BRaIn. In addition, GenLoc localizes the highest number of unique bugs (12), as shown in Fig. \ref{overlap-llm}, indicating that it complements prior LLM-based IRBL techniques. 

Notably, GenLoc successfully localizes the bug shown in Fig. \ref{rocketmq}, whereas both BRaIn and LLM-BL fail because the buggy file is never included in their initial shortlisted candidates. In contrast, GenLoc’s semantic retrieval captures the relevant file and its iterative analysis enables deeper inspection of the file’s logic. We observe similar results on the Ye et al. subset, with detailed results provided in the replication package.

\subsection{RQ3: Comparison with Issue Localization Techniques}

We evaluate GenLoc against two state-of-the-art LLM-based issue localization techniques, CoSIL and LocAgent. We use the SWE-bench Lite dataset, which is a standard benchmark adopted by recent issue localization approaches \cite{yu2025orcaloca,jiang2025cosil,chen2025locagent}. 

Table \ref{issue-comparison} presents the results, showing that GenLoc achieves the strongest performance across all evaluation metrics. It attains the highest Accuracy@1 of 60.67\%, representing a relative improvement of approximately 15\% over CoSIL and more than 110\% over LocAgent. In terms of ranking quality, GenLoc obtains MAP@10 and MRR@10 values of 0.70, yielding relative improvements of about 13\% over CoSIL and over 110\% compared to LocAgent for both metrics.

\begin{table}[h]
\footnotesize
\caption{Comparison of GenLoc and Issue Localization Techniques}
\begin{tabular}{cccccc}
\cline{1-6}
\multirow{2}{*}{\textbf{Technique}} & \multicolumn{3}{c}{\textbf{Accuracy@k (\%)}}                                                            & \multicolumn{1}{c}{\multirow{2}{*}{\textbf{MAP@10}}} & \multicolumn{1}{c}{\multirow{2}{*}{\textbf{MRR@10}}} \\ \cline{2-4}
                                    & \multicolumn{1}{c}{\textbf{k=1}} & \multicolumn{1}{c}{\textbf{k=5}} & \multicolumn{1}{c}{\textbf{k=10}} & \multicolumn{1}{c}{}                                 & \multicolumn{1}{c}{}                                 \\ \hline
GenLoc                              & \textbf{60.67}                   & \textbf{80.22}                   & \textbf{84.00}                    & \textbf{0.70}                                        & \textbf{0.70}                                        \\
CoSIL                               & 52.67                            & 74.44                            & 78.89                             & 0.62                                                 & 0.62                                                 \\
LocAgent                            & 28.33                            & 40.11                            & 41.00                             & 0.33                                                 & 0.33                                                 \\
\bottomrule
\end{tabular}
\label{issue-comparison}
\end{table}

\begin{wrapfigure}{r}{0.36\textwidth}
\vspace{-10pt}
\begin{center}
\includegraphics[width=0.36\textwidth]{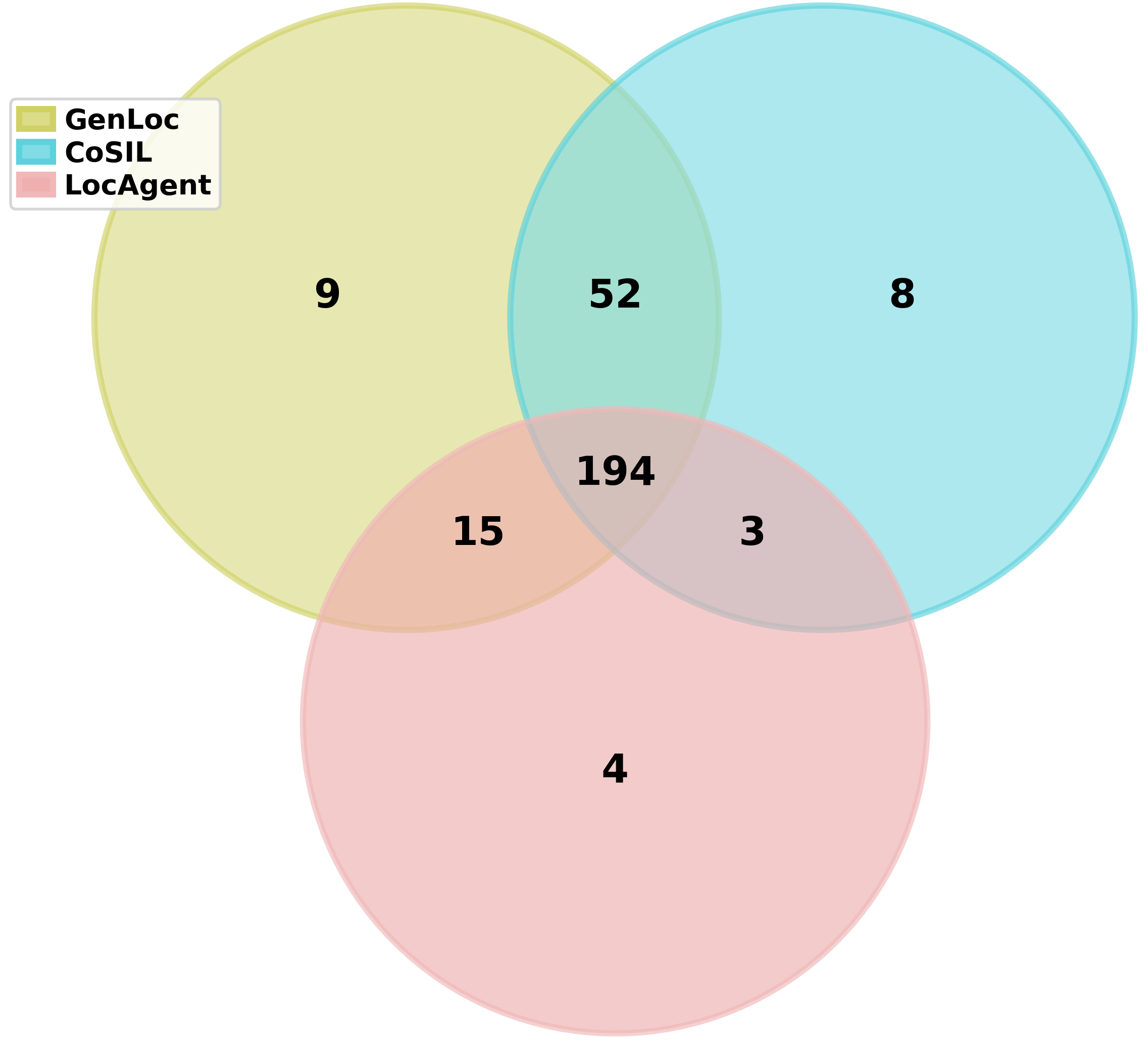}
\end{center}
\caption{Overlap Analysis between GenLoc and Other Issue Localization Approaches.}
\label{overlap-issue}
\vspace{-16pt}
\end{wrapfigure}

These results indicate that GenLoc is more effective at identifying the correct file at early ranks while maintaining a high-quality overall ranking. Furthermore, GenLoc localizes 9 unique bugs (Fig. \ref{overlap-issue}) including the bug in Fig.~\ref{pylint}. The bug describes a linting failure due to incorrect handling of directories that contain same-named module files. In this case, CoSIL is distracted by excessive repository-level context, while LocAgent is misled by lexical cues toward naming and import checkers. In contrast, GenLoc first applies semantic retrieval to narrow the search space, where {\small\texttt{expand\_modules.py}} aligned with the bug context. It then analyzes the method signatures of this file to confirm its relevance and inspects method bodies to uncover the failure-inducing logic.

\subsection{RQ4: Ablation Study}
\label{sec:rq4}
We conduct an ablation study to isolate the impact of candidate retrieval, LLM-driven code exporation with tool support and post-processing on GenLoc’s overall bug localization performance. We focus on the GHRB dataset to enable systematic evaluation of multiple GenLoc variants without prohibitive computational cost.

\subsubsection{Comparison of Syntactic and Semantic Retrieval for Shortlisting Suspicious Files}

GenLoc shortlists 50 candidate files through an initial retrieval step. To assess the effectiveness of this step independently of downstream reasoning, we compare semantic (embedding-based) retrieval with widely used BM25-based syntactic retrieval (adopted from \cite{Brain2025}) based on their ability to include the ground-truth buggy file within the top-$K$ candidates. 
Table~\ref{tab:retrieval_ablation} shows that semantic retrieval achieves a higher Accuracy@50 (83.21\%) than syntactic retrieval (71.76\%), indicating superior effectiveness in capturing bug-report intent when lexical overlap is limited. This behavior is illustrated in Fig.~\ref{rocketmq}, where syntactic retrieval fails due to limited lexical overlap between the bug report and the source code, while semantic retrieval succeeded. Thus, GenLoc adopts semantic retrieval for candidate shortlisting.

\begin{table}[ht]
\footnotesize
\centering
\caption{Comparison of Syntactic and Semantic Retrieval}
\begin{tabular}{ccccccc}
\hline
\multirow{2}{*}{\textbf{Type}} &
\multicolumn{4}{c}{\textbf{Accuracy@k (\%)}} &
\multirow{2}{*}{\textbf{MAP@10}} &
\multirow{2}{*}{\textbf{MRR@10}} \\
\cline{2-5}
 & \textbf{k=1} & \textbf{k=5} & \textbf{k=10} & \textbf{k=50} &  &  \\
\hline
Semantic Retrieval  & \textbf{20.36} & \textbf{58.27} & \textbf{72.52} & \textbf{83.21} & \textbf{0.30} & \textbf{0.37} \\
Syntactic Retrieval & 15.27 & 40.46 & 51.15 & 71.76 & 0.20 & 0.26 \\
\hline
\end{tabular}
\label{tab:retrieval_ablation}
\end{table}

\subsubsection{Impact of LLM-based Code Exploration}
To quantify the contribution of LLM-based code exploration, we compare GenLoc against several ablated variants in which key components of the pipeline are selectively disabled:

\begin{enumerate}[leftmargin=*]
\item \textbf{GenLoc:} This is the complete version of the proposed approach, where the LLM is provided with five functions, including {\small\texttt{get\_candidate\_filenames()}} to utilize embedding-based retrieval to identify potentially buggy files.

\item
\textbf{Embedding-Only/Semantic Retrieval:} A baseline that evaluates the standalone performance of the embedding-based retrieval system, where source files are ranked solely based on their semantic similarity to the bug report, without any LLM reasoning or iterative refinement.

\item \textbf{GenLoc-NoEmbed:} A variant where the LLM operates without access to the embedding-based recommended files, i.e., the {\small\texttt{get\_candidate\_filenames()}} function is removed. The model must infer filenames directly from the bug report and continue reasoning with the remaining four functions.

\item \textbf{GenLoc-Naive:} A minimal setup, where the LLM is limited to using only two functions, {\small\texttt{search\_file()}} and {\small\texttt{search\_method()}}, and must rely solely on basic keyword-based exploration without access to embedding guidance or deeper code-level analysis.

\end{enumerate}


\begin{table}[ht]
\footnotesize
\centering
\caption{Impact of Reasoning and Tooling}
\begin{tabular}{ccccccc}
\hline
\multirow{2}{*}{\textbf{Technique}} & \multicolumn{3}{c}{\textbf{Accuracy@k (\%)}}                                                            & \multicolumn{1}{c}{\multirow{2}{*}{\textbf{MAP@10}}} & \multicolumn{1}{c}{\multirow{2}{*}{\textbf{MRR@10}}} \\ \cline{2-4}
                                    & \multicolumn{1}{c}{\textbf{k=1}} & \multicolumn{1}{c}{\textbf{k=5}} & \multicolumn{1}{c}{\textbf{k=10}} & \multicolumn{1}{c}{}                                 & \multicolumn{1}{c}{}                                 \\ \hline
GenLoc                              & \textbf{63.36}                   & \textbf{76.85}                   & \textbf{79.65}                    & \textbf{0.59}                                        & \textbf{0.69}                                        \\
Embedding-Only                      & 20.36                            & 58.27                            & 72.52                             & 0.30                                                 & 0.37                                                 \\
GenLoc-NoEmbed  & 46.05                            & 56.24                            & 57.76                             & 0.42                                                 & 0.50                                                 \\ 
GenLoc-Naive  & 46.31                            & 57.25                            & 58.78                             & 0.43                                                 & 0.50                                                 \\
\hline
\end{tabular}
\label{ablation}
\end{table}

Table \ref{ablation} presents the average results on the GHRB dataset across all four configurations. Results show that GenLoc achieves the best performance on all metrics, demonstrating the strength of combining semantic retrieval with iterative LLM analysis. The Embedding-Only variant obtains only 20.36\% at Accuracy@1, indicating that retrieval alone is insufficient without deeper analysis. Similarly, the GenLoc-NoEmbed achieves only 46.05\% Accuracy@1, showing that the absence of semantic retrieval limits the LLM’s ability to identify relevant files. These results highlight that semantic retrieval and LLM-driven code analysis are complementary, and removing either degrades localization performance.


\begin{figure}[hbt]
\centering
\begin{tcolorbox}[
  top=2pt,
  bottom=2pt,
  left=2pt,
  right=2pt,
  colback=gray!10!white, 
  colframe=black!75!black, 
]
\raggedright
\scriptsize{
\textbf{Bug ID:} 20109 \\
\textbf{Summary:} [BUG] [typescript-fetch] ModelNamePrefix is added twice to import statements thereby breaking the build\\
\textbf{Description:} When generating the model files the prefix is properly added once in front of the generated files. However, in the import statements the prefix is added \_twice\_ resulting in erroneous statements breaking the build. When e.g. `SomePrefix` is set for `modelNamePrefix`, then the following model files are generated for `example-for-file-naming-option.yaml`: * `SomePrefixPetCategory.ts` * `SomePrefixPet.ts` However, \_within\_ `SomePrefixPet.ts` it looks like this: ```typescript import type { SomePrefixPetCategory } from './SomePrefixSomePrefixPetCategory';``` I.e. `SomePrefix` is added twice.}
\end{tcolorbox}
\caption{Bug Report from OpenAPI Generator Project.}
\label{openapi}
\end{figure}

The performance improvement of GenLoc over GenLoc-NoEmbed and GenLoc-Naive indicates that GenLoc goes beyond superficial cues (e.g., filenames or keywords) and performs deeper reasoning. A representative case\footnote{\url{https://github.com/OpenAPITools/openapi-generator/issues/19039}} is shown in Fig. \ref{openapi}, where GenLoc-NoEmbed and GenLoc-Naive failed to localize the correct file, while GenLoc succeeded. The bug report described an error in the TypeScript Fetch generator of OpenAPI, where the modelNamePrefix option was being incorrectly applied twice in generated import statements. Misled by surface-level clues, GenLoc-NoEmbed and GenLoc-Naive attempted to locate a nonexistent file named {\small\texttt{ModelNamePrefix}} and also searched for the {\small\texttt{generate()}} method, which was unrelated to the bug and provided no useful signal. In contrast, GenLoc invoked {\small\texttt{get\_candidate\_filenames()}} to narrow down to the TypeScript-specific implementation and by analyzing method signatures and bodies in {\small\texttt{TypeScriptFetchClientCodegen.java}} (e.g., {\small\texttt{toModelFilename()}}, {\small\texttt{parseImports()}}), it successfully localized the buggy file.

\begin{wrapfigure}{r}{0.36\textwidth}
\begin{center}
\includegraphics[width=0.36\textwidth]{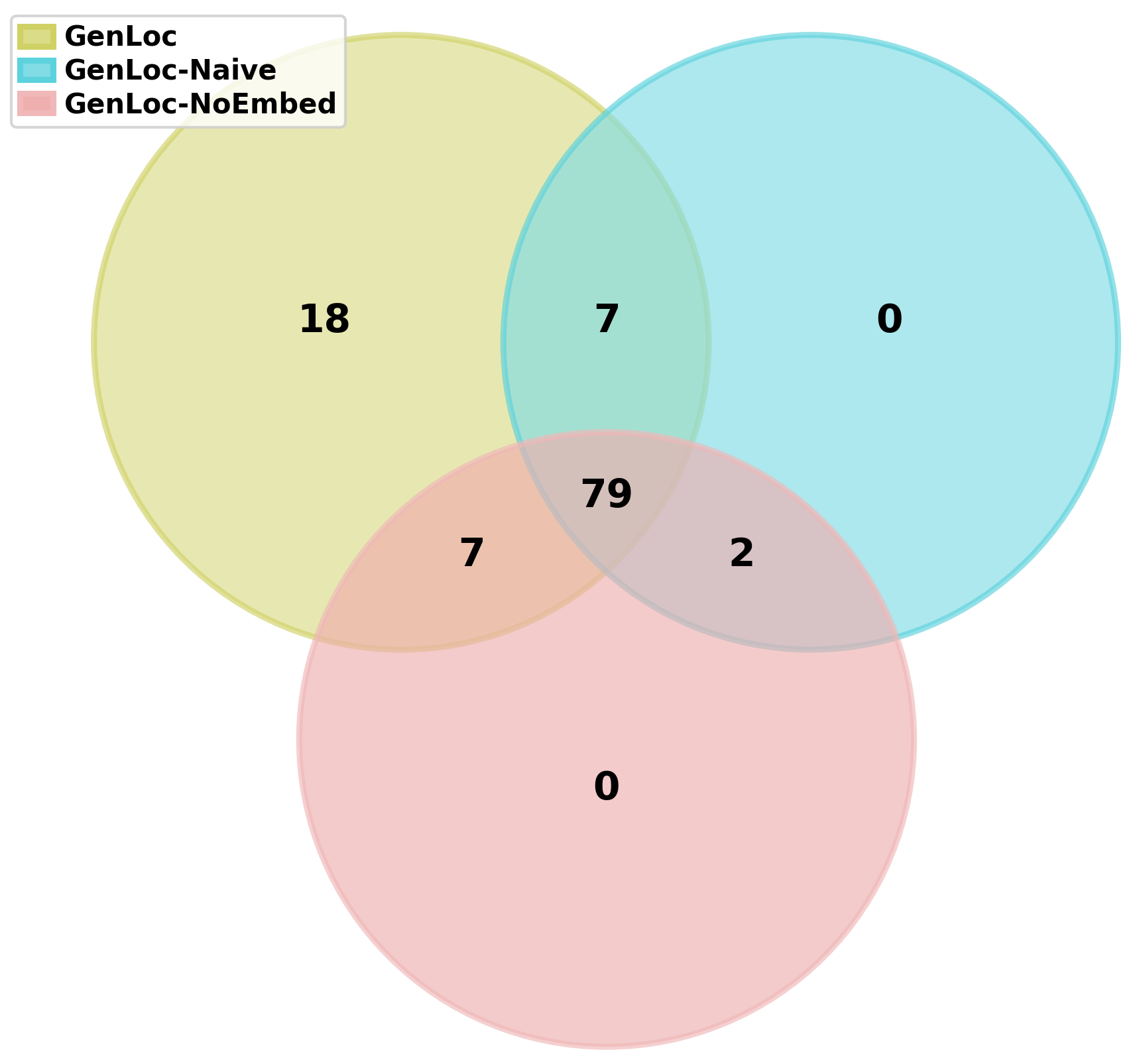}
\end{center}
\caption{Overlap analysis between GenLoc and Its Ablated Variants.}
\label{overlap-ablation}
\vspace{-35pt}
\end{wrapfigure}

Although GenLoc-NoEmbed and GenLoc-Naive obtain similar results, the overlap analysis (Fig. \ref{overlap-ablation}) shows that they capture distinct subsets of bugs. This demonstrates the importance of combining all four functions, {\small\texttt{search\_file()}}, {\small\texttt{search\_method()}}, {\small\texttt{get\_method\_signatures\_of\_a\_file()}} and {\small\texttt{get\_method\_body()}}, to achieve broader coverage. Their integration enables GenLoc to leverage both keyword-level and structure-aware reasoning.

Finally, Table~\ref{consistency} presents the consistency of results, measured as consensus across trials for each variant (i.e., GenLoc, GenLoc-NoEmbed and GenLoc-Naive). Here, consensus indicates the proportion of bugs that are localized in at least two or all three trials, reflecting the stability of each approach. GenLoc shows the highest stability, with 93.69\% of bugs localized in at least two runs and 88.29\% in all three runs. In contrast, GenLoc-NoEmbed and GenLoc-Naive attain lower values, reflecting greater variability across trials. It shows that the integration of all components not only improves accuracy but also enhances the consistency of GenLoc’s results.

\begin{table}[h]
\footnotesize
\caption{Consistency Across Trials}
\begin{tabular}{|c|c|c|c|}
\hline
\textbf{Consensus} & \textbf{GenLoc (\%)} & \textbf{GenLoc-NoEmbed (\%)} & \textbf{GenLoc-Naive (\%)} \\ \hline
$\geq$ 2 trials          & \textbf{93.69}           & 85.23                   & 86.36                 \\ \hline
$\geq$ 3 trials          & \textbf{88.29}           & 72.73                   & 76.14                 \\ \hline
\end{tabular}
\label{consistency}
\end{table}

\subsubsection{Impact of Post-Processing}
We evaluate the effect of post-processing by comparing GenLoc with a variant that disables this step (GenLoc-NoPost). Table~\ref{ablation-pp} shows that removing post-processing leads to a substantial performance drop across all metrics, primarily due to LLM's error in reconstructing fully qualified filenames (e.g., omitting intermediate package segments). Post-processing corrects these errors and refines the final ranking, leading to more accurate localization results.

\begin{table}[h]
\footnotesize
\centering
\caption{Impact of Post-Processing}
\begin{tabular}{ccccccc}
\hline
\multirow{2}{*}{\textbf{Technique}} & \multicolumn{3}{c}{\textbf{Accuracy@k (\%)}}                                                            & \multicolumn{1}{c}{\multirow{2}{*}{\textbf{MAP@10}}} & \multicolumn{1}{c}{\multirow{2}{*}{\textbf{MRR@10}}} \\ \cline{2-4}
                                    & \multicolumn{1}{c}{\textbf{k=1}} & \multicolumn{1}{c}{\textbf{k=5}} & \multicolumn{1}{c}{\textbf{k=10}} & \multicolumn{1}{c}{}                                 & \multicolumn{1}{c}{}                                 \\ \hline
GenLoc                              & \textbf{63.36}                   & \textbf{76.85}                   & \textbf{79.65}                    & \textbf{0.59}                                        & \textbf{0.69}                                        \\
GenLoc-NoPost  & 44.27                            & 49.87                            & 51.66                             & 0.47                                                 & 0.52                                                 \\ 
\hline
\end{tabular}
\label{ablation-pp}
\end{table}
\vspace{-4pt}

\subsection{RQ5: Performance on Unseen Bugs}
\label{sec:rq5}
Table~\ref{unseen} presents the performance of GenLoc on both seen and unseen bug reports from the GHRB dataset. We define unseen bugs as the subset of GHRB bug reports submitted after the knowledge cut-off date of GPT-4o-mini (October 1, 2023), resulting in 49 unseen bug reports. The results indicate that GenLoc maintains consistent performance across the two subsets. For instance, the Accuracy@1 remains stable at approximately 63\% for both seen and unseen bugs. To further validate this observation, we performed a Mann–Whitney U test, a non-parametric method suitable for comparing independent groups \cite{mcknight2010mann}. The results reveal no statistically significant difference in Reciprocal Rank (RR) (p-value > 0.7) and Average Precision (AP) (p-value > 0.3) between the two groups. These findings suggest that GenLoc does not rely on memorization or training data leakage and can effectively localize bugs even in reports that fall outside the temporal scope of the LLM's training data.

\begin{table}[ht]
\footnotesize
\centering
\caption{Performance on Seen and Unseen Bugs}
\begin{tabular}{cccccc}
\hline
\multirow{2}{*}{\textbf{Category}} & \multicolumn{3}{c}{\textbf{Accuracy@k (\%)}}                                                            & \multicolumn{1}{c}{\multirow{2}{*}{\textbf{MAP@10}}} & \multicolumn{1}{c}{\multirow{2}{*}{\textbf{MRR@10}}} \\ \cline{2-4}
                                   & \multicolumn{1}{c}{\textbf{k=1}} & \multicolumn{1}{c}{\textbf{k=5}} & \multicolumn{1}{c}{\textbf{k=10}} & \multicolumn{1}{c}{}                                 & \multicolumn{1}{c}{}                                 \\ \hline
Seen                               & 63.41                            & 78.05                            & 80.90                             & 0.60                                                 & 0.70                                                 \\
\multicolumn{1}{l}{Unseen}         & 63.27                            & 74.83                            & 77.55                             & 0.56                                                 & 0.68                                                 \\ \hline
\end{tabular}
\label{unseen}
\end{table}
\vspace{-4pt}

\subsection{RQ6: Choice of Models}
\label{sec:rq6}
We investigate the impact of both embedding models and LLM on GenLoc's performance. For embeddings, we consider OpenAI's {\small\texttt{text-embedding-3-small}} and {\small\texttt{CodeXEmbed}} (400M). Text-embedding-small is selected as a cost-effective and widely used general-purpose embedding model, while CodeXEmbed is included because it is designed to unify code and text retrieval within a single framework and has demonstrated strong performance on code retrieval task \cite{liu2024codexembed}. As for LLMs, we compare {\small\texttt{GPT-4o-mini}} with {\small\texttt{GPT-5.1}}. GPT-4o-mini represents a lightweight and cost-effective model, whereas GPT-5.1 is one of the latest and more advanced LLMs. Similar to \hyperref[sec:rq4]{RQ4}, we focus on the GHRB dataset to enable systematic evaluation of multiple GenLoc variants without prohibitive computational cost.

Table~\ref{model-choice} reports the corresponding results. Both embedding models yield comparable performance across all metrics, indicating that GenLoc is relatively robust to the choice of embedding model as long as the embeddings effectively capture both code and natural language semantics. In contrast, the choice of LLM reveals a clear accuracy–cost trade-off. GPT-5.1 consistently achieves higher localization accuracy than GPT-4o-mini, but incurs a per-bug cost that is approximately 6 times higher (about \$0.035 vs. \$0.006 per bug). Despite this cost difference, GPT-4o-mini delivers competitive performance, making it suitable for large-scale or budget-constrained settings, whereas GPT-5.1 is preferable when maximizing localization accuracy is the primary objective.


\begin{table}[ht]
\footnotesize
\centering
\caption{Performance of Different Models}
\begin{tabular}{llccccc}
\hline
\multirow{2}{*}{\textbf{Embedding}} &
\multirow{2}{*}{\textbf{LLM}} &
\multicolumn{3}{c}{\textbf{Accuracy@k (\%)}} &
\multirow{2}{*}{\textbf{MAP@10}} &
\multirow{2}{*}{\textbf{MRR@10}} \\
\cline{3-5}
 &  & \textbf{k=1} & \textbf{k=5} & \textbf{k=10} &  &  \\
\hline
\multirow{2}{*}{Text-embedding-small}
 & GPT-4o-mini & 63.36 & 76.85 & 79.65 & 0.59 & 0.69 \\
 & GPT-5.1     & \textbf{70.74} & \textbf{85.50} & \textbf{88.04} & \textbf{0.66} & \textbf{0.77} \\
\hline
\multirow{2}{*}{CodeXEmbed}
 & GPT-4o-mini & 60.82 & 77.86 & 80.41 & 0.58 & 0.68 \\
 & GPT-5.1     & \textbf{69.21} & \textbf{86.51} & \textbf{90.08} & \textbf{0.66} & \textbf{0.77} \\
\hline
\end{tabular}
\label{model-choice}
\end{table}
\vspace{-4pt}

\section{Threats to Validity}
This section presents potential ways in which the validity of the study may be compromised.

\smallskip
\noindent
\textbf{Internal Validity:} One of the most significant threats to internal validity is the possibility of data leakage, such as bug-fixing commits being present in the LLM's training data. However, the result of \hyperref[sec:rq5]{RQ5} shows that GenLoc maintains consistent performance on both seen and unseen bug reports in the GHRB dataset, indicating that GenLoc effectively localizes bugs even for reports outside the temporal scope of the LLM's training data. Another potential threat arises from the inherent randomness of LLMs, which may yield different results across different runs. To mitigate this, all experiments were repeated three times, and a replication package is released to support reproducibility and enable further validation by the research community.


\smallskip
\noindent \textbf{Construct Validity:} Construct validity concerns whether the evaluation metrics accurately measure the effectiveness of bug localization techniques \cite{widyasari2022influence}. This study employs Accuracy@k, MRR@10 and MAP@10, which are widely used in prior IRBL research and considered standard for assessing ranking quality \cite{wong2023handbook,rahman2018improving}.

\smallskip
\noindent \textbf{External Validity:} GenLoc is evaluated on three complementary benchmarks spanning two programming languages. The first is a large benchmark of over 9,000 bug reports from six large open-source Java projects that has been widely used in prior IRBL research \cite{chakraborty2024rlocator}. The second is the GHRB dataset which contains 131 recent bugs from 16 Java projects and is specifically designed for evaluating LLM-based IRBL approaches \cite{lee2024github}. The third is SWE-bench Lite, which consists of 300 issues from 12 popular Python projects \cite{xia2025agentless}. While these benchmarks provide diverse evaluation settings, the findings may still not generalize to proprietary industrial systems or projects written in other programming languages. Additional studies are needed to evaluate the effectiveness of GenLoc in broader and more heterogeneous settings.

\section{Conclusion and Future Work}
This paper presents GenLoc, a novel IRBL technique that integrates semantic retrieval with LLM-driven iterative code analysis. We evaluate GenLoc on over 9,000 real-world bugs from six large-scale Java projects, the GHRB dataset and the SWE-Bench Lite dataset. Across all datasets, GenLoc consistently outperforms traditional IR-based methods, deep learning models, state-of-the-art LLM-based IRBL techniques and the file-level localization stages of recent issue localization approaches in terms of Accuracy@k, MRR@10 and MAP@10. Moreover Genloc can localize the largest number of unique bugs that other techniques fail to identify, demonstrating its effectiveness over existing IRBL techniques and issue localization pipelines.

In the future, GenLoc could be extended to support finer-grained bug localization at the method or statement level to provide more precise debugging assistance. Another promising direction is to integrate GenLoc with automated program repair pipelines to evaluate its effectiveness in guiding correct patch generation.

\section{Data Availability}
Our replication package\footnote{\url{https://github.com/mou23/Towards-Explorative-IRBL}} is publicly available to facilitate future research and reproducibility. 











\bibliographystyle{ACM-Reference-Format}
\bibliography{arXiv}


\end{document}